\documentclass[aps,twocolumn,showpacs,amsmath,amssymb,superscriptaddress,final,10pt,prb]{revtex4-1}

\usepackage{graphicx}
\usepackage{color}
\usepackage{dcolumn} 
\usepackage{bm}      
\usepackage{nicefrac}
\usepackage{mathrsfs}

\newcommand{\BACSO}{BaCu$_2$Si$_2$O$_7$}

\begin{document}

\title{Dimensional crossover of spin chains in a transverse staggered field: an NMR study}

\author{F.\ Casola}
\email{fcasola@phys.ethz.ch}
\affiliation{Laboratorium f\"ur Festk\"orperphysik, ETH H\"onggerberg, CH-8093 Z\"urich, Switzerland}
\affiliation{Paul Scherrer Institut, CH-5232 Villigen PSI, Switzerland}

\author{T.\ Shiroka}
\affiliation{Laboratorium f\"ur Festk\"orperphysik, ETH H\"onggerberg, CH-8093 Z\"urich, Switzerland}
\affiliation{Paul Scherrer Institut, CH-5232 Villigen PSI, Switzerland}

\author{V.\ Glazkov}
\affiliation{P.L. Kapitza Institute for Physical Problems, RAS, 119334 Moscow, Russia}

\author{A.\ Feiguin}
\affiliation{Department of Physics, Northeastern University, Boston, Massachusetts 02115, USA}

\author{G.\ Dhalenne}
\affiliation{Laboratoire de Physico-Chimie de l'Etat Solide, Universit\'e Paris-Sud, 91405 Orsay cedex, France}

\author{A.\ Revcolevschi}
\affiliation{Laboratoire de Physico-Chimie de l'Etat Solide, Universit\'e Paris-Sud, 91405 Orsay cedex, France}

\author{A.\ Zheludev}
\affiliation{Neutron Scattering and Magnetism, Laboratorium f\"ur Festk\"orperphysik, ETH H\"onggerberg, CH-8093 Z\"urich, Switzerland}

\author{H.-R.\ Ott}
\affiliation{Laboratorium f\"ur Festk\"orperphysik, ETH H\"onggerberg, CH-8093 Z\"urich, Switzerland}

\author{J.\ Mesot}
\affiliation{Laboratorium f\"ur Festk\"orperphysik, ETH H\"onggerberg, CH-8093 Z\"urich, Switzerland}
\affiliation{Paul Scherrer Institut, CH-5232 Villigen PSI, Switzerland}

\date{\today}

\begin{abstract}
Heisenberg spin-\nicefrac{1}{2} chain materials are known to substantially alter their static and dynamic properties when experiencing an effective transverse staggered field originating from the varying local environment of the individual spins. We present a temperature-, angular- and field-dependent $^{29}$Si NMR study of the model compound \BACSO. The experimental data are interpreted in terms of the divergent low-temperature transverse susceptibility, predicted by theory for spin chains in coexisting longitudinal and transverse staggered fields. Our analysis first employs a finite-temperature ``Density Matrix Renormalization Group" (DMRG) study of the relevant one-dimensional Hamiltonian. Next we compare our numerical with the presently known analytical results. With an analysis based on crystal symmetries we show how the anisotropic contribution to the sample magnetization is experimentally accessible even below the ordering temperature, in spite of its competition with the collinear order parameter of the antiferromagnetic phase. The modification of static and dynamic properties of the system due to the presence of a local transverse staggered field (LTSF) acting on the one-dimensional spin array are argued to cause the unusual spin reorientation transitions observed in \BACSO. On the basis of a Ginzburg-Landau type analysis, we discuss aspects of competing spin structures in the presence of magnetic order and the enhanced transverse susceptibility.
\end{abstract}

\pacs{75.10.Pq, 76.60.-k, 75.40.Cx}


\maketitle

\section{\label{Intro}Introduction}
In systems of reduced dimensionality, thermal and quantum fluctuations are the source of physical phenomena without analogues in ordinary 3D materials. Since the original studies of Mermin and Wagner on the role of thermal fluctuations,\cite{MW66} it is known that in the isotropic spin-$S$ Heisenberg model the divergent number of low-energy thermal excitations (i.e.\ spin waves) suppresses any long-range ordered phase for 
dimensions $d \leq 2$.
 Due to the reduced dimensionality, the magnetic properties of spin arrays such as spin chains and ladders are further influenced by quantum effects that are masked in common 3D-materials.
At $T=0$, despite quantum corrections present in the antiferromagnetic case, magnetic order develops in the ground state of the 2D bipartite Heisenberg model. Due to the statistical analogy of a quantum $d$-dimensional antiferromagnet at zero temperature with a purely classical magnet at finite temperature in $d+1$ dimensions,\cite{SpinChains02} quantum effects, particularly effective in the antiferromagnetic case, imply the absence of long-range order in quantum Heisenberg antiferromagnets for $d=1$.

In real bulk materials, however, the same concept of one-dimensionality is ill-defined. For instance, it has often been shown that the effective dimensionality of certain materials can strongly change upon variations of temperature,\cite{Lake05} magnetic field,\cite{Kramer07} or energy scale\cite{Zheludev01,Lake09} being probed. As a consequence, the interpretation of the experimental observations requires an interpolation between limits of different effective degrees of freedom.
This is particularly true for the specific case of spin-\nicefrac{1}{2} antiferromagnetic (AF) Heisenberg chains realized in well-known model materials such as KCuF$_3$,\cite{Lake05} Sr$_2$CuO$_3$,\cite{Kojima97} or copper pyrazine dinitrate.\cite{Stone03,Reich11}
Their high-temperature disordered phases, well described by the strictly one-dimensional Luttinger Liquid concept\cite{GiamarchiBook} are, at low temperatures, replaced by a magnetically ordered state induced by weak interchain interactions. The need to interpolate between different dimensionalities suggested the introduction of the class of so-called \emph{quasi}-1D materials. The degree of one-dimensionality is usually measured in terms of the ratio between the N\'eel temperature, marking the onset of the 3D magnetic order, and the intrachain exchange interaction.\cite{Broholm02,Zheludev00Hal,Kenz01} Values of this ratio much below one signify the preservation of the 1D character of the system. Nevertheless, 1D quantum fluctuations 
continue to be effective in the 3D domain too, i.e., below the N\'eel temperature.
This is evident from both the strongly reduced saturation moment in the ordered state as well as from the deviations with respect to excitations predicted by the standard spin-wave theory.\cite{Zheludev02,Lake99} Observations of the evolution of the properties of a physical system across the phase diagram, in regions characterized by different effective dimensionalities, has motivated a number of theoretical and experimental studies, dedicated to the phenomenon of dimensional crossover.

The physics of an assembly of isolated spin-\nicefrac{1}{2} quantum chains upon increasing the interchain couplings is rather well understood.\cite{Zheludev02,Fabian97} The impact of such a perturbation on chains experiencing a local transverse staggered field (LTSF), however, is still an open question.\cite{Xi11} A spin-\nicefrac{1}{2} quantum chain in a uniform field and 
a concomitant LTSF is usually modeled by the Hamiltonian:\cite{Oshikawa97}
\begin{equation}
\mathscr{H} = J \sum_{i} \mathbf{S}_i \cdot \mathbf{S}_{i+1} - g \mu_{\mathrm{B}} H \sum_{i} S^z_i + (-1)^i \mu_{\mathrm{B}} H_{\perp} \sum_i S^x_i \label{Hamorig} 
\end{equation}
where $i$ is the site-index along the chain direction, $J$ is the intrachain exchange coupling constant, $\mu_{\mathrm{B}}$ is the Bohr magneton, $g$ is the electronic gyromagnetic ratio, $z$ and $x$ are the directions parallel and perpendicular to the externally applied field $H$, respectively. The locally induced transverse staggered field is $H_{\perp} = c H$, with $c$ as a constant.

The interest in spin chains with an LTSF was first triggered by experimental studies on copper benzoate Cu(C$_6$H$_5$COO)$_2$$\cdot$3H$_2$O,\cite{Date70} probing the compound's magnetism via susceptibility, nuclear magnetic resonance (NMR) and electron-spin resonance (ESR) measurements. In zero magnetic field ($H=0$), copper benzoate was known to be just another realization of a spin-\nicefrac{1}{2} Heisenberg chain, with an exchange constant $J/k_{\mathrm{B}} \sim 18$~K and the onset of a canted antiferromagnetic order at $\sim 0.8$~K.\cite{Dender96} 
Due to the relatively small intrachain exchange coupling, copper benzoate was considered as a favorable material for studying the field-dependent properties of the 1D spin-\nicefrac{1}{2} quantum Heisenberg model. However, the field-dependent bulk properties were soon found to be incompatible with corresponding theoretical results.\cite{Oshima76}
Incommensurate soft modes at field-dependent reciprocal space positions of the excitation spectrum are expected for spin-\nicefrac{1}{2} antiferromagnets in an applied field.
The first inelastic neutron scattering (INS) experiment, intended to verify these theoretical predictions, was performed on copper benzoate. \cite{Dender97} Field-driven incommensurate modes were indeed found at the expected positions in reciprocal space. Due to an unexpected field- and orientation-dependent spin-gap $\Delta$, scaling as $H^{\nicefrac{2}{3}}$, these modes were not soft, however.\cite{Dender97}
This surprising INS observation was soon interpreted as the result of the particular character of the $g$-tensors and Dzyaloshinskii-Moriya (DM) interactions. Along the chain direction, the non-diagonal tensor components change sign from one Cu site to the next and the same is true for the DM $\mathbf{D}$-vector.
Both can be mapped onto an effective LTSF and thus to Eq.~\eqref{Hamorig}.
Formally it turns out that, in the absence of interchain perturbative terms, the effective low-energy theory of the LTSF Hamiltonian is given by the quantum sine-Gordon (SG) model.\cite{Affleck99,Essler99} The SG model is one of the few non-linear problems which benefits from an exact solution. Hence, the low-energy theory provides analytical expressions for the physical quantities in the temperature range $\Delta <T \ll J$ when $g \mu_{\mathrm{B}} H \ll J$. \\

In the last 15 years the magnetic properties of a large number of both organic and inorganic compounds have successfully been described by the LTSF model. Specifically, NMR and ESR studies focused mostly on the joint detection of the temperature-dependent longitudinal ($\propto \langle S^z_i \rangle (T)$) and transverse ($\propto \langle S^x_i \rangle (T)$) local magnetization. Examples are the cases of copper pyrimidine dinitrate \cite{Wolter05,Feyerherm00} and of BaCu$_2$Ge$_2$O$_7$ \cite{Bertaina04}.

Quite generally, any real material, unless directly excluded by symmetry arguments, will develop an arbitrarily small LTSF, captured in the $H_{\perp}$ term of Eq.~\eqref{Hamorig}, when an external field $H$ is applied. This field-induced LTSF is expected to compete with that developing due to interchain interactions below the N\'eel temperature ($T<T_{\mathrm{N}}$). The resulting physics in the limit where both effects are of similar magnitude is currently not clear.\cite{Xi11} One of the few experimental studies where both perturbations on the 1D spin-\nicefrac{1}{2} Heisenberg model were strong enough to be experimentally visible is an INS experiment on CuCl$_2$ $\cdot$ 2(dimethylsulfoxide) (CDC).\cite{Kenzelmann04} Spin excitations measured at 40 mK showed that the competition leads to the opening of a gap at a non zero value of the staggered field $H_{\perp}$, rather than in zero field, as the scaling relation $\Delta \propto H^{\nicefrac{2}{3}}$ would imply. Besides the particular case of CDC, all other ESR and NMR studies of materials modeled by Eq.~\eqref{Hamorig} were usually limited to the range $T>T_{\mathrm{N}}$. 

By exploiting the high sensitivity of nuclear magnetic resonance to local magnetic fields, we present a $^{29}$Si NMR study of \BACSO. We demonstrate how the local magnetization in a chain, adequately modeled by Eq.~\eqref{Hamorig}, develops above $T_{\mathrm{N}}$ and how it evolves when $T \leq T_{\mathrm{N}}$. The latter evolution could be monitored thanks to the complete decoupling of the local field-induced magnetization from the order parameter which is related to the onset of a standard, temperature-driven second-order magnetic phase transition.

In section \ref{sec:PreviousRes} we briefly summarize relevant results of previous research on \BACSO. Our NMR data are presented in section \ref{sec:results}, while a detailed discussion of the analysis is reported in section \ref{DataAnaly}. Finally, section \ref{sec:summ} offers a summary and the main conclusions of this work.

\section{B\lowercase{a}C\lowercase{u}$_2$S\lowercase{i}$_2$O$_7$: Summary of previous research}
\label{sec:PreviousRes}
\begin{figure}
\includegraphics[width=0.5\textwidth]{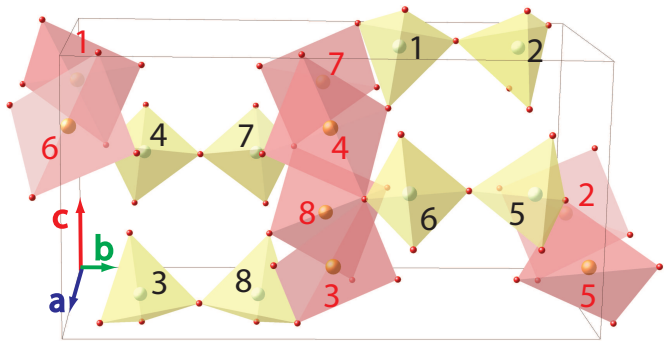} 
\caption{\label{fig:Figure1} a) Schematic view of the crystalline unit cell of \BACSO . Red and black numbers identify the copper- and silicon atom sites, respectively. The small red spheres indicate the oxygen sites. Zigzag Cu$^{2+}$ spin chains run along the $c$ axis.}
\end{figure}

\BACSO\ crystallizes in the orthorhombic space group $Pnma$ ($D^{16}_{2h}$) with lattice constants $a=6.862$ \AA, $b=13.178$ \AA\ and $c=6.897$ \AA.\cite{Yamada00} 
Each unit cell contains eight Cu$^{2+}$ ions, which determine the material's magnetic properties, and eight silicon atoms, each family being equivalent by local symmetry. Both types of atoms, together with the respective closest-oxygen-atom configurations, are depicted in Fig.~\ref{fig:Figure1}. The copper-ion spins form zigzag chains along the crystallographic $c$ axis.
Early zero-field INS studies of \BACSO\ at $T>T_{\mathrm{N}}$ revealed a gapless spinon continuum. A fit to the lower boundary of the spectrum provided an intrachain exchange value $J_1=24.1$ meV.\cite{Tsukada99}
 At the same time, elastic neutron-scattering measurements found evidence of a long-range antiferromagnetic order below $T_{\mathrm{N}} = 9.2$ K, with an ordered Cu$^{2+}$ moment at saturation of 0.15~$\mu_{\mathrm{B}}$, collinear with the $c$ axis.\cite{Tsukada99,Kenz01} These results qualify \BACSO\ to be among the best realizations of a spin-$\nicefrac{1}{2}$ Heisenberg chain system, characterized by a low ordering temperature  and by a large intrachain exchange coupling. For a global overview of a material classification the reader is referred to Table 1 in Ref.~\onlinecite{Broholm02}.
Over the years, studies of \BACSO\ proceeded along two main directions, both related to the present work: \emph{i}\/) the study of the spin-reorientation transitions \cite{Tsukada01,Zhed02,Glazkov05,Poirier02} and, \emph{ii}\/) the investigation of the 1D-to-3D crossover.\cite{CrossPRLs}

Regarding the first topic, experiments employing neutron scattering,\cite{Tsukada01,Zhed02} ESR,\cite{Glazkov05} and ultrasound techniques,\cite{Poirier02} were carried out in externally applied magnetic fields of varying strengths and orientations. These studies provided evidence of a number of phase transitions related to the realignment of spins (and spin-flop transitions for an externally applied field $\mathbf{H} \parallel c$) at $T<T_{\mathrm{N}}$.\cite{Glazkov11} In spite of serious efforts, the cause for these transitions is not yet clear.\cite{Glazkov05,Glazkov11}
In the following we show that NMR data provide direct evidence of the presence of an enhanced 
transverse susceptibility below $T_{\mathrm{N}}$, offering a simple explanation for the observed spin reorientations. We suggest that non-trivial phase diagrams below the magnetic ordering temperature may appear naturally in anisotropic quasi-1D antiferromagnets exhibiting a strong reduction of the ordered moment.

As for the second topic, early studies have shown that \BACSO\ is also a model spin-chain compound for investigating the crossover from quantum-spin 1D dynamics to semi-classical 3D (spin-wave) dynamics.\cite{Zheludev01,Kenz01} Zero-field INS experiments probing the range below the N\'eel temperature revealed how the presence of weak interchain interactions $J_2 \ll J_1$, produce an effective LTSF originating from neighboring chains, which leads to a confinement of massless spinon excitations at $T<T_{\mathrm{N}}$. Therefore the situation is not equivalent to that of non interacting 1D chains in a staggered field. The additional $J_2$ interactions induce dispersive excitations with a momentum transverse to the chain direction. Their energy vanishes at well defined positions in reciprocal space, coinciding with the Bragg peaks of the ordered structure. 
For \BACSO\ in an applied magnetic field, intrinsic anisotropies lead to a field-induced LTSF causing a spin-gap to be present already above $T_{\mathrm{N}}$. The possibility of directly measuring the field-induced gap via INS is slim, but important consequences are expected for the local static magnetization. Since the latter is accessible via NMR, we took advantage of this unique opportunity to study the 1D-to-3D crossover in a quantum-spin chain with an LTSF modeled by Eq.~\eqref{Hamorig}.

\section{ NMR Experimental Results}  
\label{sec:results}
The NMR measurements were carried out on a $4\times2\times2$ mm$^3$ single crystal of \BACSO. Since the $a$ and $c$ lattice parameters are roughly the same, the correct identification of the two crystallographic directions is not trivial. Field-dependent magnetization measurements on the same sample with $\mathbf{H} \parallel c$ could detect the two expected spin-flop transitions, hence confirming the correct identification of 
the crystal axes.

\begin{figure*}
\includegraphics[width=1\textwidth]{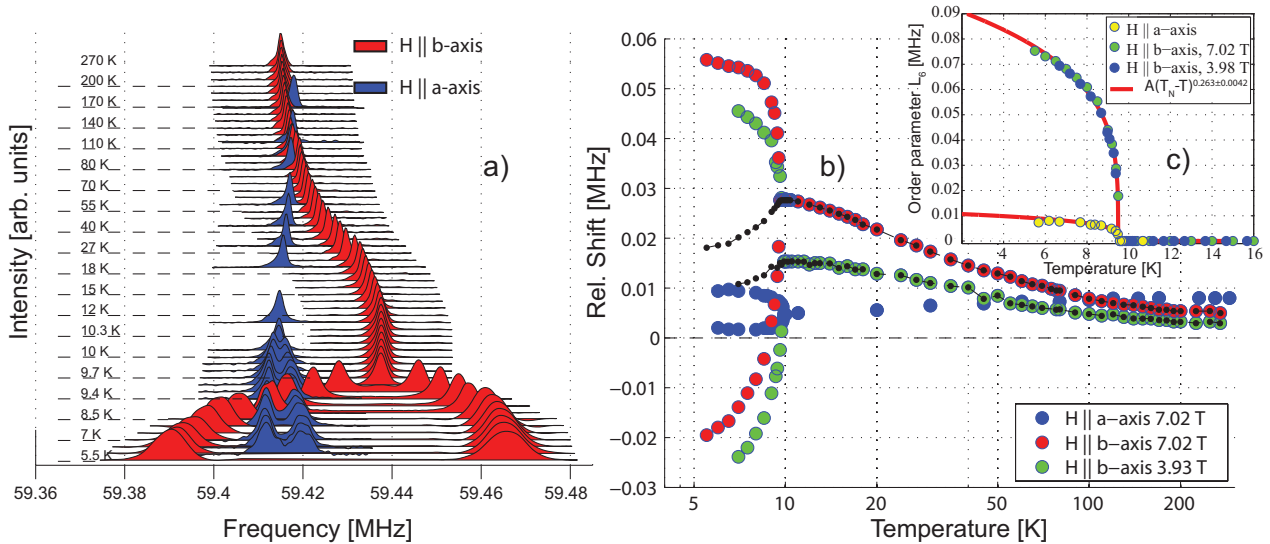} 
\caption{\label{fig:Figure2} a) Temperature dependence of NMR $^{29}$Si resonances with an applied field $\mathbf{H}$ of 7.02 T along the crystallographic $b$ and $a$-axes. b) Relative shift of the peak positions as a function of temperature. c) Direct measure of the order parameter of the antiferromagnetic phase, reflected in the difference of the positions of the split lines. The mean values in the ordered phase are indicated by black dots. The zero frequency marks the resonance frequency of $^{29}$Si nuclear spins in a standard reference sample such as Si(CH$_3$)$_4$.\cite{HARRIS01}}
\end{figure*}

The $^{29}$Si NMR lines  were measured as a function of temperature at two different magnetic fields, with the external field being applied along the crystallographic $a$ or $b$ axis. Spectra above the antiferromagnetic transition were recorded with a standard spin--echo technique, while at lower temperatures the line intensities were established by a superposition of frequency sweeps as described in Ref.~\onlinecite{Clark95}.    
The NMR spectra reported in Fig.~\ref{fig:Figure2}a show the height-normalized shapes vs.\ temperature, measured at 7 T for both crystal orientations.
In the paramagnetic phase ($T > T_{\mathrm{N}}$) single lines are observed, which split into two in the ordered regime. While the results are qualitatively the same for both field directions, quantitative differences are obvious. With the field along the $b$ axis, the NMR line shifts towards higher frequencies (by ca.\ 30 kHz) upon lowering the temperature. A much smaller shift is observed when $\mathbf{H} \parallel a$ and likewise the line splitting for $T < T_{\mathrm{N}}$ is considerably reduced. The signal width, instead, is approximately the same in both cases and is essentially unaffected accross the entire temperature range in the paramagnetic regime.

The monotonic decrease of the resonance frequency with increasing temperature 
in the paramagnetic phase (see Fig.~\ref{fig:Figure2}b) is quite surprising. A spin-$\nicefrac{1}{2}$ Heisenberg chain is expected to display a broad maximum in the magnetic susceptibility $\chi(T)$ at $T_{\mathrm{max}} \simeq 0.64 J$.\cite{Klumper00}
This maximum, also known as the Bonner-Fisher peak,\cite{Bonner64} 
should be located at $T_{\mathrm{max}} \simeq 180$ K for \BACSO\ ($J = 24.1$ meV). 
This was indeed reported in previous works\cite{Tsukada01} and confirmed by our low-field magnetization data (see Fig.~\ref{fig:FigureCOMP}c). Since the NMR line shift and the magnetic susceptibility are generally proportional, the absence of any maximum in Fig.~\ref{fig:Figure2}b for $T > T_{\mathrm{N}}$ is a striking feature, indicating a fundamental difference between the microscopically probed local field (via NMR) and that reflected in the macroscopic susceptibility.

Once the lineshape maxima in the 3D ordered regime ($T < T_{\mathrm{N}}$) were identified (see Fig.~\ref{fig:Figure2}b), the differences between the peak positions, $\Delta \nu$, were evaluated. The corresponding values for the two different orientations are shown in Fig.~\ref{fig:Figure2}c. We recall that $\Delta \nu$ is ultimately proportional to the order parameter of the phase transition and hence it can be used to monitor the transition. The resulting increase in frequency splitting between 10 and 5 K can be fitted by a power-law $A \cdot \left( T_{\mathrm{N}} - T \right)^{\beta}$. Although the chosen temperature range is too broad to really reflect a truly critical regime, the exponent $\beta = 0.263\pm 0.004$ is very close to $\beta = 0.25$, the value obtained from zero-field neutron diffraction data.\cite{Kenz01} 
Figure~\ref{fig:Figure2}c also confirms that the monotonic increase of the order parameter for $T < T_{\mathrm{N}}$ does not depend on field. 
Indeed, for $\mathbf{H} \parallel b$, practically the same ordered moment at saturation is found for $\mu_0 H = 3.98$ T and 7 T. 

The postulated collinear antiferromagnetic order for $T < T_{\mathrm{N}}$ is known to have a zero-field saturation moment of 0.15 $\mu_{\mathrm{B}}$ and an easy axis which coincides with the $c$ direction.\cite{Tsukada99,Kenz01}
As shown in Fig.~\ref{fig:Figure2}, the line positions reflect the NMR response to a magnetic field oriented along the $a$ or $b$ direction, respectively, i.e., perpendicular to the easy axis $c$.
A standard collinear antiferromagnet with the field applied perpendicular to the easy axis exhibits a constant magnetization in the ordered phase. The NMR lines are thus supposed to split symmetrically with respect to their common relative shift,\cite{Yosida} clearly at variance with our observation. In fact, the average positions of the two maxima, indicated in Fig.~\ref{fig:Figure2}b by full black dots,
contrary to expectations, are observed to decrease with decreasing temperature for $T < T_{\mathrm{N}}$. Spin-wave corrections to the constant magnetic susceptibility below $T_{\mathrm{N}}$ for antiferromagnets ordered collinearly along a direction perpendicular to the applied field show at most an increase of the longitudinal magnetization with decreasing temperature. This is due to zero-point spin fluctuations affecting the ordered moment.\cite{Jongh10} However, according to the data presented in Fig. \ref{fig:FigureCOMP}c, this correction is modest in \BACSO. 

\begin{figure}
\includegraphics[width=0.4\textwidth]{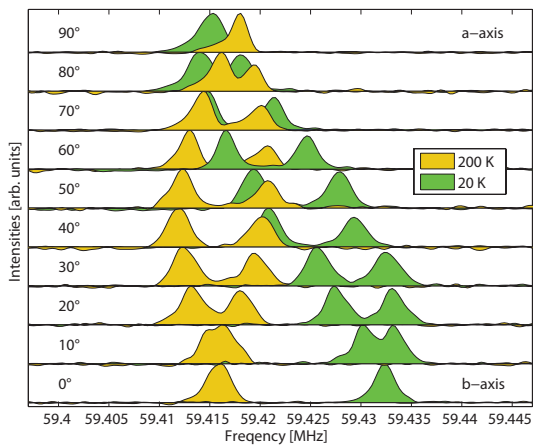} 
\caption{\label{fig:Figure3} Angular dependence of the NMR lines at 20 and 200 K. The orientation of the external field $\mu_0 H = $7.02 T varies within the crystalline $ab$ plane.}
\end{figure}

The ${}^{29}$Si NMR resonances of \BACSO\ depend strongly on sample orientation, a trend which is particularly conspicuous in the ordered phase, where the positions and shapes of the NMR lines are sensitive to even a small degree of misalignment. In order to avoid problems with data interpretation due to misorientation, a study to establish the orientation dependence of the NMR lines was carried out by mounting the sample on a two-axis goniometer, suitable for NMR experiments at cryogenic temperatures.\cite{NMRgoniom} The results for two temperatures above $T_{\mathrm{N}}$, 20 and 200 K, are reported in Fig.~\ref{fig:Figure3}. Once the $a$ and $b$ axes were identified, the sample was rotated such that the direction of the externally applied field was kept in the $ab$ plane of the crystal lattice.

\section{Data Analysis}
\label{DataAnaly}
\subsection{Origin of the staggered field in \BACSO}
\label{sec:LTSF_origin}
As already mentioned in Sec.~\ref{Intro}, the zigzag geometry of the Cu$^{2+}$ spin chains in \BACSO\ provokes electron anisotropies which may strongly affect the physics of the chain system in case of an externally applied magnetic field. Two dominant contributions to the anisotropy originate either in off-diagonal components of the gyromagnetic tensor $g$ (alternating in sign along the chain direction), and/or in spin-orbit effects in the Cu-O-Cu superexchange path along the chain. The latter is also known as the Dzyaloshinskii-Moriya (DM) interaction.\cite{DMref} 
To compare the LTSF model of Eq.~\eqref{Hamorig} with experimental data, an estimate of these two contributions to anisotropy has to be made. 
A gap in the spin-wave excitation spectrum in the ordered phase was observed in zero-field INS measurements and attributed to two-ion anisotropy effects with an energy scale of $\sim 0.4$ meV,\cite{Kenz01} while an additional mode at a lower energy of $\sim$ 0.17 meV was observed by ESR.\cite{Glazkov05} Based on the local symmetry of the intrachain Cu-O-Cu bond, a DM $\mathbf{D}$-vector lying almost in the $ab$ plane, with unit vector components [0.86, 0.51, 0.07] was suggested.\cite{Tsukada01}

\begin{figure}
\includegraphics[width=0.45\textwidth]{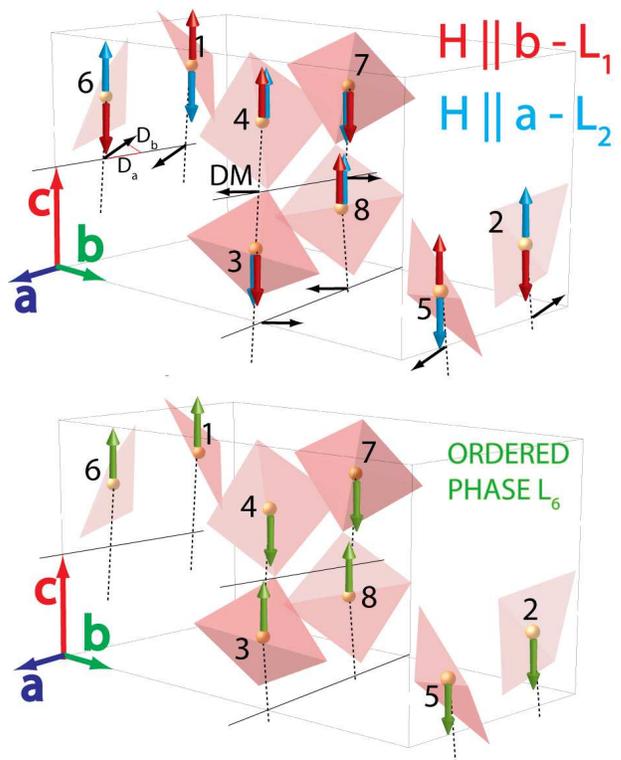} 
\caption{\label{fig:Figure4} Upper panel: Pattern of the field-induced LTSF in \BACSO. Red (blue) lines correspond to the case of a longitudinal field applied along the $b$($a$)-axis. The LTSF orientation of sites 3,4,7 and 8 is not affected by the change of field orientation. Black arrows show the direction of the DM vector according to the symmetry rule in Eq.~\eqref{DMsymm} (see text). Lower panel: Local magnetization pattern of the order parameter in the magnetically ordered phase at $T<T_N$ and $\mathbf{H}$= 0. The same kind of structure is realized with a moderate external field in the $ab$-plane.}
\end{figure}

By making use of the crystal symmetry, we apply general space-group operations $\{R_{\alpha},\mathbf{\tau}_{\alpha}\}$ to a pair $ij$ of copper sites interacting via oxygen superexchange along the $c$-axis. Rotations $R_{\alpha}$ and affine transformations $\tau_{\alpha}$ are related to the symmetry operation $\alpha$. The original pair is transformed into a new set and the local environment transforms accordingly.
The configuration of the various $\mathbf{D}$ vectors in the unit cell can be established from the transformation rule:\cite{Yosida} 
\begin{equation}
\{R_{\alpha},\mathbf{\tau_{\alpha}}\} \mathbf{D}_{i,j} = \mathbf{D}_{\{R_{\alpha},\mathbf{\tau}_{\alpha}\}i,\{R_{\alpha},\mathbf{\tau}_{\alpha}\}j}, \label{DMsymm}
\end{equation}
where $\mathbf{D}$ transforms as an axial vector under the application of the rotation $R_{\alpha}$. With this rule the full pattern of alternating $\mathbf{D}$ vectors, depicted as black arrows, halfway between the relevant Cu sites in the upper panel of Fig.~\ref{fig:Figure4}, can be derived. 
We note that the $a$ and $b$ components of the $\mathbf{D}$ vector have alternating signs when moving along a given chain in the $c$ direction, or when moving between different chains in the $a$ direction.\footnote{For instance, if we assume $\mathbf{D}_{3,4} = (D_a, D_b, D_c)$ we obtain $\mathbf{D}_{4,3'} = (-D_a, -D_b, D_c) = \mathbf{D}_{8,7}$, $\mathbf{D}_{3,4} = \mathbf{D}_{7,8'}$ and also $\mathbf{D}_{2,1} = (D_a, -D_b, -D_c) = \mathbf{D}_{6,5'}$, $\mathbf{D}_{5,6} = (-D_a, D_b, -D_c) = \mathbf{D}_{1,2'}$. The prime after the index site denotes copper ions belonging to the upper nearest-neighbor unit cell along the $c$ direction.}
The effect of the DM interactions between sites $i$ and $j$, if not forbidden by crystal symmetry, can be taken into account via the following spin Hamiltonian:\cite{DMref}
\begin{equation}
\mathscr{H}_{\mathrm{DM}} = \mathbf{D}_{i,j} \cdot \left( \mathbf{S}_i \times \mathbf{S}_j \right). \label{Dmpart}
\end{equation}
This contribution can be mapped onto a local transverse staggered field, $\mathbf{H}^i_{\perp}$, via a rotation in spin space.\cite{Affleck99} 
For small $|\mathbf{D}|/J$ ratios, as is the case for \BACSO, the local transverse field at site $i$ can be approximated by:
\begin{equation}
\mathbf{H_{\perp}}^i \simeq  \frac{1}{2J} \mathbf{D}_{j,i} \times g_i^u \mathbf{H}, \label{mappingDM}
\end{equation}
where $g_i^u$ is the uniform (diagonal) part of the local $g$-tensor valid at site $i$ and $\mathbf{H}$ is the externally applied field. The term in Eq.~\eqref{mappingDM} represents the second type of the LTSF components outlined above. 

In addition, strongly orientation-dependent, high-temperature magnetization data were interpreted as indicating a strong anisotropy of the local $g$-tensor.\cite{Tsukada01} 
A direct measurement of the $g$-tensor components is in general possible via ESR experiment. In the case of \BACSO, however, this is hampered by the broadening and the loss of intensity of the ESR absorption in the paramagnetic phase, providing at best an estimation of $g$-factor $g_b=2.11\pm0.07$ and $g_c=2.0\pm0.1$.\cite{Glazkov05} The evaluation of the local $g$-tensor is additionally complicated by the strong in-chain exchange interaction, leading to the exchange (or motional) narrowing of the resonance line.\cite{Anderson54} Consequently, differences in the $g$-factor cannot be resolved as long as the corresponding Zeeman splittings are smaller than the exchange energy. 
Since $J/k_{\mathrm{B}}\simeq 200$ K, this is clearly the case here, even in very strong fields. As a working hypothesis we assume that the $g$-factor anisotropy is mostly determined by single-ion effects, which reflect the local configuration of oxygen atoms. Figure~\ref{fig:Figure1} shows that each copper ion is located at the center of a tetrahedrally distorted CuO$_4$ square. Because of this distortion, the oxygen-to-oxygen distances of opposite O atoms differ by about 2\%. If we neglect this detail, the local point group of the CuO$_4$ unit is $D_{2d}$. An arbitrary $g$-tensor is then invariant under all the symmetry operations of the point group $D_{2d}$. In the local reference frame of a CuO$_4$ unit, the tensor adopts a uniaxial form ${g}^{\mu,\nu}$$= \mathrm{diag}(\tilde{g}_1$, $\tilde{g}_1$, $\tilde{g}_9$), with two of the principal axes oriented along the two directions at 45 degrees from the square's diagonals and the third axis parallel to their cross product. 
The transformation matrix relating a CuO$_4$ unit to the crystallographic unit cell may now be obtained. 
Selecting the copper atom 1 in Fig.~\ref{fig:Figure1} this transformation leads to:
\begin{equation}
{g}_1^{\mu,\nu}=
\begin{bmatrix}   
0.23 \tilde{g}_1 +  0.7 \tilde{g}_9 & 0.31(\tilde{g}_1-\tilde{g}_9) & 0.33(\tilde{g}_1-\tilde{g}_9) \\
0.31(\tilde{g}_1-\tilde{g}_9) & 0.86\tilde{g}_1+0.14\tilde{g}_9 & 0.15(\tilde{g}_9-\tilde{g}_1) \\
0.33(\tilde{g}_1-\tilde{g}_9) & 0.15(\tilde{g}_9-\tilde{g}_1) & 0.84\tilde{g}_1+0.15\tilde{g}_9
\end{bmatrix},
\label{gFinal}
\end{equation}
where the subscript $i$ of ${g}_i^{\mu,\nu}$ refers to the site index. Since the eight copper sites are equivalent under the allowed symmetry operations, we can obtain the tensor ${g}^{\mu,\nu}$ for each of them. The matrix in Eq. \eqref{gFinal} reveals that the components $g_i^{2,2}$ and $g_i^{3,3}$ are roughly equal (in qualitative agreement with ESR experiments), reflecting the coinciding high-temperature magnetization tails, measured with a field along the $b$ and $c$ direction, respectively. The qualitative behavior of the magnetization is also shown in Fig.~1 of Ref.~\onlinecite{Tsukada01}. 

By considering the $g$-tensor for each site $i$ in the unit cell, and by using Eq.~\eqref{mappingDM} for the DM contribution to the local field, the expected LTSF pattern for an external field applied along the $a$ or $b$ axis can be established. This pattern is the key for understanding the NMR results. In our case, whenever the external field lies in the $ab$ plane, the LTSF $H^i_{\perp}$ is found to be parallel to the $c$ direction (see Fig. \ref{fig:Figure4}). In order to exploit this favorable configuration, the field orientation $\mathbf{H} \parallel c$ has not been addressed in the present work. 
With the external field in the $ab$ plane and forming an angle $\theta$ with the b-axis, we can calculate the ratio $H^i_{\perp}/H$.  
For instance, by considering the copper sites 3 and 5 we obtain:
\begin{align} 
c^3 = \frac{H^3_{\perp}}{H} & = \left(-g_1^{3,2}\cos{\theta}+g_1^{3,1}\sin{\theta} \right) + \nonumber \\
& \frac{1}{2J}\left(-\text{D}_ag_1^{2,2}\cos{\theta} - \text{D}_bg_1^{1,1}\sin{\theta} \right); \nonumber \\
c^5 = \frac{H^5_{\perp}}{H} & = \left(g_1^{3,2}\cos{\theta}+g_1^{3,1}\sin{\theta} \right) + \nonumber \\
& \frac{1}{2J}\left(\text{D}_ag_1^{2,2}\cos{\theta} - \text{D}_bg_1^{1,1}\sin{\theta} \right). \label{LTSF1}
\end{align}  
The configurations of the LTSF, resulting from the orientation of an external field along the crystalline $b$ ($a$) direction, 
are indicated by red (blue) arrows in the upper panel of Fig.~\ref{fig:Figure4}. The terms $c^i$ in \eqref{LTSF1} are to be inserted in Eq. \eqref{Hamorig}.
These two patterns are found to be equivalent to the $L_{1c}$ and $L_{2c}$ irreducible representations of the magnetic-structure space, as previously established in Ref.~\onlinecite{Glazkov04}. They consist of the following linear combinations:
\begin{align} 
L_{1c} = S_{1c} - S_{2c} - S_{3c} + S_{4c} + S_{5c} - S_{6c} - S_{7c} + S_{8c}, \nonumber \\
L_{2c} = S_{1c} - S_{2c} + S_{3c} - S_{4c} + S_{5c} - S_{6c} + S_{7c} - S_{8c}, \label{CombinationsLz}
\end{align}  
with $S_{ic}$ the component of the local magnetization at the $i$-site along the $c$-axis.
Note that the products $L_{1c}H_b$ and $L_{2c}H_a$ are symmetry invariants,\cite{Balan17} i.e., they are combinations of the irreducible representations (IR) of the magnetic structure which transform according to the trivial representation (the 1D representation consisting of 
$1 \times 1$ matrices containing the entry 1) of the little group of the propagation vector $k$. \footnote{This is the case when either the space group is symmorphic, or the propagation vector is $k=0$.}

\subsection{Static magnetization of a spin-\nicefrac{1}{2} chain in a transverse staggered field}
\label{sec:LTSF_physics}

As mentioned above the local magnetization at a copper site $i$, denoted as $\mathbf{S}^i $, is given by a uniform and a transverse component, such that $\mathbf{S}^i = \mathbf{S}^i_{u} + \mathbf{S}^i_{\perp}$, locally induced by an external uniform field $\mathbf{H}_{u} = g^u_i \mathbf{H}$ and a staggered field  $\mathbf{H}^i_{\perp}$. In the following we fix the convention that $\mathbf{S}^i$ has a saturation value of \nicefrac{1}{2}. In order to recall the general results already known for $\mathbf{S}^i_{u}$ and $\mathbf{S}^i_{\perp}$, and to present our new results, it is useful to introduce the following reduced units:
\begin{align}
&h^*_{u} = \frac{g_{u} \mu_{\mathrm{B}} H}{J}, \qquad  \chi^*_{u} = \frac{\partial \langle S_{u} \rangle}{\partial h^*_{u}}, \nonumber \\
&h^*_{\perp,i} = \frac{\mu_{\mathrm{B}} H^i_{\perp}}{J}, \qquad \chi^*_{\perp,i} = \frac{\partial \langle S^i_{\perp} \rangle}{\partial h^*_{\perp,i}}, \nonumber \\
& T^* = \frac{T}{J}. \label{Renormaliz}
\end{align}

\begin{figure}
\includegraphics[width=0.5\textwidth]{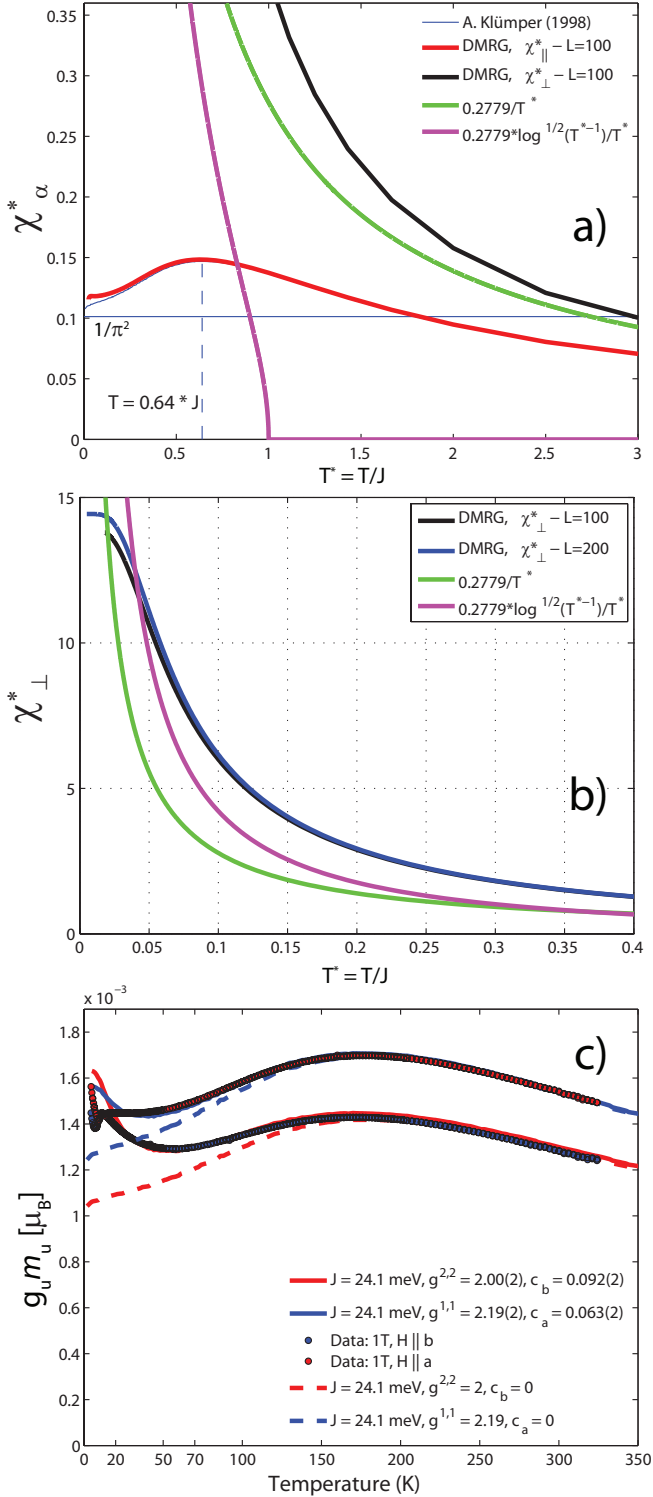} 
\caption{\label{fig:FigureCOMP} a) Results of finite temperature DMRG calculations across an extended range of $T^*$ based on Hamiltonian \eqref{Hamorig} and valid in the small $h^*_{\perp}$ limit, in comparison with known analytical and tabulated results (see text for details). b) Comparison of the calculated $\chi^*_{\perp}$ as obtained analytically or via DMRG at $T^* < 0.4$. c) SQUID-magnetometry data measured at 1 T with the field applied along the $a$ and $b$ crystalline axes, respectively. Solid lines are fits using Eq.~\eqref{Feiguin_renormalized}, dashed lines are predictions for an isotropic uniform chain with no LTSF.}
\end{figure}
It has been shown \cite{Johnston00,Klumper98} that by using this notation and setting both $h^*_{\perp,i}$ and $h^*_{u}=0$,  the susceptibility $\chi^*_{u}$ has a peak at $T^*\simeq0.64$ and a zero-temperature limit of $1/\pi^2$. In Fig.~\ref{fig:FigureCOMP}a we reproduce the temperature dependence of $\chi^*_{u}$, calculated and tabulated by A. Kl\"{u}mper in Ref.~\onlinecite{Klumper98}. The transverse field $h^*_{\perp}$ opens a gap in the excitation spectrum which, for $h^*_{\perp} \ll 1$, scales as:\cite{Shibata01}
\begin{equation}
\frac{\Delta}{J} = 1.78 \cdot \left( h^*_{\perp} \right)^{\frac{2}{3}} \cdot \left( -\log^{\nicefrac{1}{6}}  h^*_{\perp} \right) 
\label{gapLTSF}
\end{equation}
For small fields and $\Delta/J \ll T^* \ll 1$ analytic field-theoretical results for $\chi^*_{\perp}$ are available.\cite{Affleck99} In the chosen reduced units it reads:
\begin{equation}
\chi^*_{\perp} = \frac{0.2779 \log^{\nicefrac{1}{2}}({T^*}^{-1}) }{T^*} \label{Affleck_pred}
\end{equation}
The high $h^*_{\perp}$ limit has been treated by previous ``Density Matrix Renormalization Group" (DMRG) calculations.\cite{Shibata01}
No complete and unbiased numerical result is yet available for the susceptibility at small magnetic fields across an extended range of $T^*$.
In Fig. \ref{fig:FigureCOMP}a,b we fill this gap  with results of DMRG calculations\cite{Feiguin05,White1993,White1992} for chains with 100 and 200 spin sites, respectively, and compare them with the analytical result of Eq.~\eqref{Affleck_pred}, including or omitting the logarithmic correction. In our figure, the subscript $\alpha$ stands for $u$ or $\perp$, respectively.
 Without the logarithmic term, the main difference between the analytical and the numerical result is, as expected, at low temperatures. In order to see how much a transverse field affects the longitudinal uniform magnetization, we followed I. Affleck's approach \cite{Affleck99} and computed numerically the total derivative $-dF/dh^*_{u} = m_{u}$ of the free energy of the system obtaining:
\begin{equation}
m_{u} = \langle S_{u} \rangle + c^i \langle S^i_{\perp} \rangle, \label{Feiguin_renormalized}
\end{equation}
with the parameters $c^i$ as given in Eq.~\eqref{LTSF1}. 
Two separate DMRG runs were employed to calculate the uniform and staggered susceptibilities independently. Subsequently equation \eqref{Feiguin_renormalized} was used to obtain the value of $m_{u}$. The possibility to compute the two quantities $\langle S_{u} \rangle$ and $\langle S^i_{\perp} \rangle$ separately, allowed us to make use of symmetries to reduce the Hilbert space in the DMRG simulations.

In order to compare our simulation results with the data for \BACSO, the temperature dependence of the magnetization $g_{u}m_{u}$ was measured along the $a$- and the $b$-axis of the crystal. The data, measured with a SQUID magnetometer, are displayed in Fig. \ref{fig:FigureCOMP}c. Because of the small moment, particular care was taken in the choice of the sample-holder material that would cause at most a small magnetic background signal.
With the model given by Eq.~\eqref{Feiguin_renormalized} we obtain good agreement with the experimental data down to approximately 20 K. The departure of the solid lines from the points is most likely due to approaching the onset of magnetic order. As previous authors\cite{Tsukada01} we also tried a fit by imposing the value $c^i=0$ (dashed lines). The resulting discrepancies are obvious. From the high-temperature tails of $m_{u}$($T$) we extract values between 2.19 and 2 for $g^{1,1}$ and $g^{2,2}$, respectively. The latter differ considerably, but are more realistic than the corresponding values between 2.5 and 2.2 quoted in Refs.~\onlinecite{Tsukada01} and ~\onlinecite{Tsukada99}. Useful information can be extracted from the fit parameters $c^i$. For the macroscopic uniform magnetization $m_{u}$, the sign change of $c^i$ (see Eq.~\ref{LTSF1}) is irrelevant. Thus in Fig.~\ref{fig:FigureCOMP}c, $c_a$ and $c_b$ are the corresponding parameters for fields $\mathbf{H}$ along the $a$- or $b$-direction, respectively. First of all we consider the field-induced and angle-dependent spin-gap in Eq. \eqref{gapLTSF}. We obtain $\Delta_a = 0.61$ meV and $\Delta_b = 0.78$ meV. These gaps, when expressed in $\Delta/k_{\mathrm{B}}$ units, are both of the order of 10 K, i.e.\ close to the temperature where the magnetic order sets in. This explains why an activated behavior of the spin-lattice relaxation time has not been observed in our previous NMR work.\cite{Toni11} A proper estimate for the DM parameters $D_{a,b}$ can be obtained by solving Eq.~\eqref{LTSF1} with ${g}_1^{\mu,\nu}$ from \eqref{gFinal} and the fitted  $c_{a,b}$. We obtain the values $D_a \simeq 0.94$ meV and $D_b \simeq -1.2$ meV. Remarkably, the ratio $D_a/D_b \simeq 1.25$ is close to 1.68, the value obtained from purely geometrical considerations.\cite{Tsukada01} 
Experimentally it turns out that the NMR response at low temperatures is dominated by the transverse magnetization $\langle S_i^{\perp} \rangle$, i.e., the diverging susceptibility $\chi^*_{\perp}$ emerging from the DMRG calculation in Fig. \ref{fig:FigureCOMP}b. The study of this quantity via NMR and its fate below $T_{\mathrm{N}}$ is the main topic of the rest of this paper.

\subsection{Modeling the NMR lines}
\label{sec:themodel}
Having established the contributions to the local magnetic field at the Cu sites, we now proceed to study their influence on the ${}^{29}$Si NMR-line data. In our case the local magnetization experienced by the silicon nuclei is dominated by the externally applied field, and the contribution due to the sample's magnetization is only of second order. For this reason the resonance frequency ${}^{29}\omega_{k}$ of the $^{29}$Si nucleus $k$ ($k = 1,\ldots, 8$ --- see Fig.~\ref{fig:Figure1} for the notation) can be written as:\cite{Vachon08}
\begin{equation} 
{}^{29}\omega_{k} \simeq \gamma \frac{\mathbf{H}}{|\mathbf{H}|} \cdot \left[ (1 + \mathbf{\sigma}_{k}) \cdot \mathbf{H} + \sum_{i=1}^{\infty} \mathsf{T}^i_{k} \cdot \mathbf{S}^i + \sum_{i=1}^{NN} \mathsf{A}^i_{k} \cdot \mathbf{S}^i   \right], \label{couplingSi}
\end{equation}
where $\gamma$ is the $^{29}$Si nuclear gyromagnetic ratio, $\mathbf{\sigma}_{k}$ is the orbital shift tensor, $\mathsf{T}^i_{k}$ is the dipolar tensor which couples the silicon nucleus $k$ to the copper atom $i$, and $\mathsf{A}^i_{k}$ is the relevant transferred hyperfine interaction.
The dipolar sum in Eq.~\eqref{couplingSi} can be calculated directly.
This was done by fixing the Cu-spin arrangement resulting from the LTSF configuration shown in Fig. \ref{fig:Figure4} and by including the contributions from the copper atoms within 50~\AA\ from the considered silicon site. 
The sum of hyperfine interactions runs over the four nearest neighbor (NN) copper sites.\cite{Toni11} 
Given $\mathsf{A}^i_{k}$ and  $\mathsf{T}^i_{k}$ for the silicon nucleus $k$, the relevant tensors for the other silicon sites can be obtained by allowed symmetry transformations, meaning that tensors referring to the various silicon nuclei are not mutually independent. If a symmetry operation of the space group  $\{R_{\alpha},\mathbf{\tau}_{\alpha}\}$, applied to the silicon site $\mathbf{r}_{k}$, brings it to $\{R_{\alpha},\mathbf{\tau}_{\alpha}\}\mathbf{r}_{k} = \mathbf{r}_{k'}$ (and consequently the copper site $\mathbf{r}_{i}$ to $\mathbf{r}_{i'}$), the hyperfine tensors are given by $\mathbf{A}^{i'}_{k'} = R^T_{\alpha} \cdot \mathbf{A}^i_{k} \cdot R_{\alpha}$. For example: 
\begin{align}
\mathbf{A}^3_{1} = \left(
\begin{array}{ccc}
a_1 & a_2 & a_3 \\
a_4 & a_5 & a_6 \\
a_7 & a_8 & a_9
\end{array} \right) \rightarrow
\mathbf{A}^5_{2} = \left(
\begin{array}{rrr}
a_1 & -a_2 & a_3 \\
-a_4 & a_5 & -a_6 \\
a_7 & -a_8 & a_9
\end{array} \right).
\label{transfa}
\end{align}
The four NN copper atoms surrounding the silicon atom located at site $k = 1$ are $i = 3, 4, 7, 8$ (see Fig.~\ref{fig:Figure1}). 
Contrary to dipolar interactions, the components of the transferred-hyperfine tensor are \emph{a priori} unknown. 
For calculating directly the $\theta$-dependent component $h^{hf}_{k, u}$ of the uniform local field at site $k = 1$ or $2$, parallel to $g^u \mathbf{H}$, we define $\mathbf{A}^3_{1} = a_{\mu}$, $\mathbf{A}^4_{1} = b_{\mu}$, $\mathbf{A}^7_{1} = c_{\mu}$ and $\mathbf{A}^8_{1} = d_{\mu}$ and we denote $m = g_{u}m_{u}$ and $m^i = g^{3,3} \langle S_i^{\perp} \rangle$.
By using the notation of Eq.~\eqref{transfa}, we obtain:

\begin{align}
h^{hf}_{k,u} &= \left( \begin{array}{rr} \sin{\theta} & \cos{\theta} \end{array} \right) \cdot
\left[  
 \left( \begin{array}{rr}
a_1 & \pm a_2 \\
\pm a_4 & a_5 \\
\end{array} \right) +
 \left( \begin{array}{rr}
b_1 & \pm b_2 \\
\pm b_4 & b_5 \\
\end{array} \right)  \right. \nonumber \\
& + \left. \left( \begin{array}{rr}
c_1 & \pm c_2 \\
\pm c_4 & c_5 \\
\end{array} \right) +
 \left( \begin{array}{rr}
d_1 & \pm d_2 \\
\pm d_4 & d_5 \\
\end{array} \right) 
\right] \cdot
\frac{m}{g_{u}}
\left( \begin{array}{c} g_1^{11} \sin{\theta} \\ g_1^{22} \cos{\theta} \end{array} \right) \nonumber \\
&=\frac{m}{2g_{u}}
\left[
Y_1 g_1^{11} + Y_5 g_1^{22} + \left( Y_5 g_1^{22} - Y_1 g_1^{11} \right) \cos{2\theta} + \right. \nonumber \\
& \left. \pm \left( Y_2 g_1^{22} + Y_4 g_1^{11} \right) \sin{2\theta}
\right],
\label{uniformPart}
\end{align}
where the plus (minus) sign refers to $k = 1$ (2). 
In \eqref{uniformPart} we take into account that the sample's uniform longitudinal magnetization and the externally applied field may not be collinear due to a possible $g$-tensor anisotropy. By definition, $g^2_{u} = (g_1^{1,1} \sin{\theta})^2+(g_1^{2,2} \cos{\theta})^2$ and $Y_{\mu} =  a_{\mu} + b_{\mu} + c_{\mu} + d_{\mu}$. 
The reason for picking the Si sites $k = 1, 2$ for describing the relevant NMR lineshapes is evident from the contribution of the transverse field $h^{hf}_{k,\perp}$ to the resonance frequency. We get:
\begin{align}
h^{hf}_{1/2,\perp} &= \left( \begin{array}{rr} \sin{\theta},  & \cos{\theta} \end{array} \right) \cdot
\left[ m^{3/5}
 \left( \begin{array}{r}
a_3 \\
\pm a_6 \\
\end{array} \right) + m^{4/6}
 \left( \begin{array}{r}
b_3\\
\pm b_6\\
\end{array} \right) \right. \nonumber \\
& + \left. m^{7/1}
 \left( \begin{array}{r}
c_3\\
\pm c_6\\
\end{array} \right) + m^{8/2}
 \left( \begin{array}{r}
d_3\\
\pm d_6 \\
\end{array} \right) 
\right]. \label{stgloc12}
\end{align}   
In the \emph{paramagnetic phase} the following relations always hold by symmetry:

\begin{align}
m^6 &= -m^5 \qquad m^1 = m^5 \qquad m^2 = -m^5 \nonumber \\
m^4 &= -m^3 \qquad m^7 = m^3 \qquad m^8 = -m^3. \label{relationsmSmall}
\end{align}
By denoting $G_{3/6} = a_{3/6}-b_{3/6}+c_{3/6}-d_{3/6}$, we obtain for the transverse local field in the paramagnetic phase:
\begin{align}
h^{hf}_{1/2,\perp} &=  m^{3/5} (G_3 \sin{\theta} \pm G_6 \cos{\theta}). \label{finalTransv}
\end{align}   
It turns out that by considering any other of the silicon sites, the only two possible \emph{orthogonal} local fields are those given by \eqref{finalTransv}. For an applied field along the $a$ or the $b$ axis these two local fields coincide. From Eq.~\eqref{LTSF1} it follows that if $\theta=0^{\circ}$ (field along $b$), $m^3 = -m^5$. The same is true if $\theta=90^{\circ}$ (field along $a$) (see Fig.~\ref{fig:Figure4}). A single narrow line is thus expected for $\theta=0^{\circ}$ and $\theta=90^{\circ}$, while two lines are expected in an intermediate angular range and at temperatures exceeding $T_{\mathrm{N}}$. This is indeed the case, as already shown in Fig.~\ref{fig:Figure3}.

We conclude this section with two principal results. The first concerns the prediction for the angular-, temperature- and field-dependent relative NMR line shift $\Delta \omega$ in the paramagnetic ($T > T_{\mathrm{N}}$) regime due to the transferred-hyperfine and orbital interactions:
\begin{align}
{}^{29}\Delta \omega_{1/2} &= \gamma \left \{ \frac{m}{2g_{u}}
\left[
Y_1 g_1^{11} + Y_5 g_1^{22} + \left( Y_5 g_1^{22} - Y_1 g_1^{11} \right) \cos{2\theta} \right. \right. \nonumber \\
& \left. \left. \pm \left( Y_2 g_1^{22} + Y_4 g_1^{11} \right) \sin{2\theta}
\right] \right. \nonumber \\
& \left. + m^{3/5} \left( G_3 \sin{\theta} \pm G_6 \cos{\theta} \right) \right. \nonumber \\
& \left. + \frac{1}{2} \left[ \sigma_1 + \sigma_5 + \left( \sigma_5 - \sigma_1 \right) \cos{2\theta} \pm 
2 \sigma_2 \sin{2\theta}
\right]
\right \},
\label{SHIFT1}
\end{align}
where a symmetric orbital-shift tensor $\sigma$ has been introduced. The consequences of Eq.~\eqref{SHIFT1} are discussed in the following section. The model behind Eq.~\eqref{SHIFT1} is ultimately independent of the exact geometry of the hyperfine couplings; the qualitative result does not change even if, for instance, the parameters $c_{\mu}$ and $d_{\mu}$ were zero.

\begin{figure*}
\includegraphics[width=0.9\textwidth]{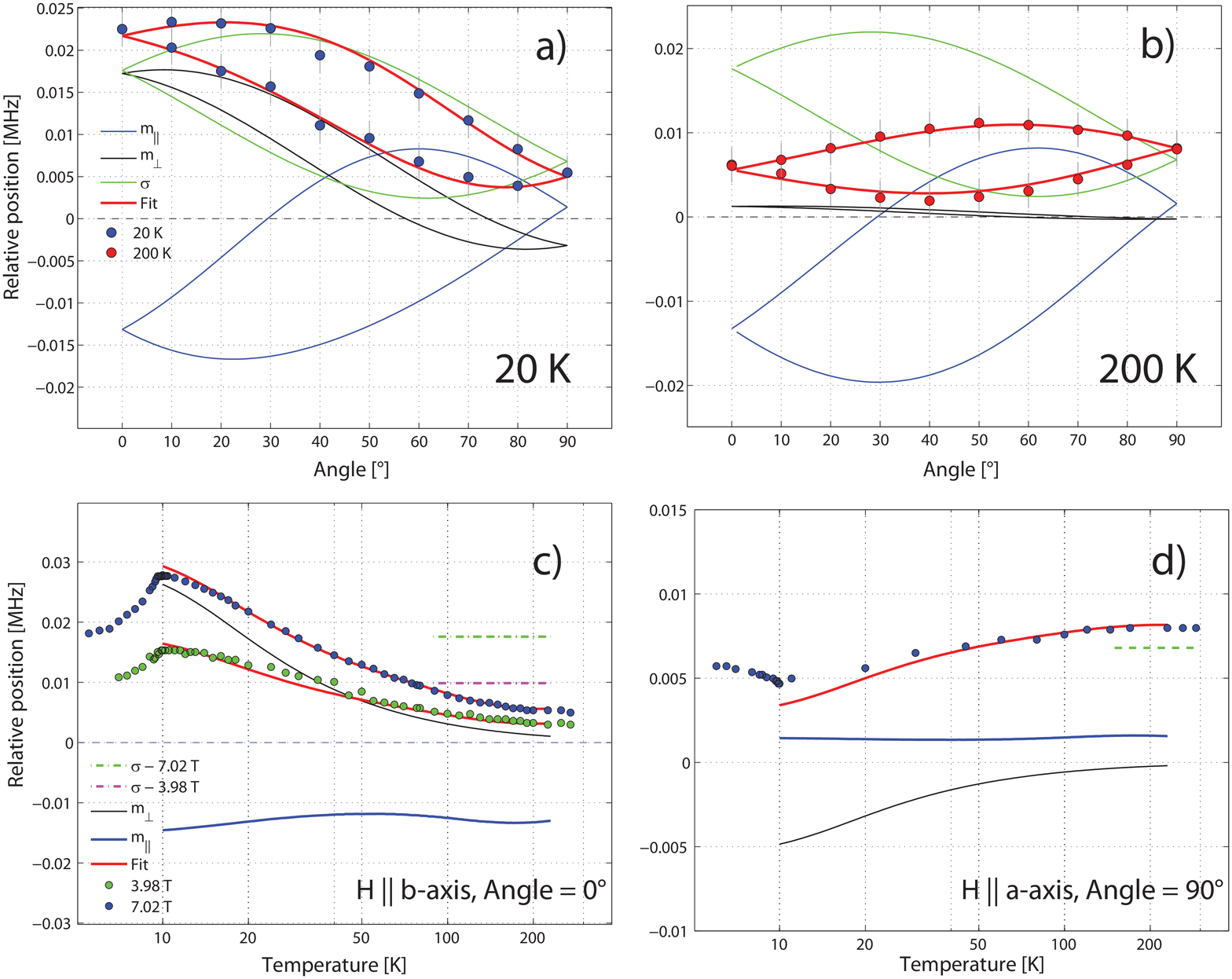} 
\caption{\label{fig:Figure5} Detailed comparison of the microscopic model captured in Eq.~\ref{SHIFT1} with the experimental $^{29}$Si NMR line positions for $T>T_{\mathrm{N}}$. In each panel the green, black and blue lines represent individual contributions to the fit related to the orbital shift, the transverse staggered and the uniform longitudinal magnetization, respectively. a,b) Angular dependence of the positions of the line maxima with $\mathbf{H}$ in the $ab$-plane, measured at 20 K and 200 K ($b$ axis corresponds to $\theta = 0$). c,d) Temperature dependence of the NMR shifts measured with the field applied along the $b$ (left) or $a$ (right) axis. The zero frequency marks the undisturbed resonance frequency of $^{29}$Si nuclear spins.}
\end{figure*}

Next we discuss the situation in the \emph{ordered regime}, below $T_{\mathrm{N}}$. Here, due to the second-order phase transition, the symmetry of the system is spontaneously broken.  The adopted order reflects one of the irreducible representations of the magnetic structure. Its product with the applied field is, however, not necessarily an invariant upon symmetry transformations. Consequently, in the ordered regime the lines are expected to split, even when the field is applied along the main crystallographic axes. From previous zero-field diffraction studies\cite{Kenz01} and following the conventions in Ref.~\onlinecite{Glazkov04}, it is known that the representation chosen by the spin system is $L_6$, collinear with the $c$ axis (see lower panel of Fig.~\ref{fig:Figure4}). 
By calculating the local field employing Eq.~\eqref{finalTransv}, the NMR lines split below $T_{\mathrm{N}}$ according to:
\begin{align}
h^{hf}_{1/2,\perp} &=  m^{3/5}_{T<T_{\mathrm{N}}} ( \tilde{G}_3 \sin{\theta} \pm \tilde{G}_6 \cos{\theta}). \label{finalTransvORD}
\end{align}   
Since in the $L_6$ representation $m^3 = -m^5$ is always valid, Eq.~\eqref{finalTransvORD} suggests that the lines should coincide if $\theta =0$. This is not the case, however, if all the possible silicon sites are considered. In the ordered phase, the local field at the sites $k=1, 2, 4, 7$ is the same. Also the sites $k = 3, 5, 6, 8$ experience the same field, but the latter differs slightly from the former. This explains the line splitting shown in Fig. \ref{fig:Figure2} for both field orientations with respect to the crystal axes $a$ and $b$.

 The parameters $\tilde{G}_{3/6} =$ $a_{3/6}-b_{3/6}-c_{3/6}+d_{3/6}$ $\neq G_{3/6}$ are not directly accessible by experiment. Nevertheless,   Eq.~\eqref{finalTransvORD} offers the possibility to average out the $L_6$ contribution to the NMR shift below $T_{\mathrm{N}}$, thus providing a \emph{\textbf{direct access}} to the components $m$ and $m^i$ of the local magnetization in the ordered regime.

\subsection{Comparison between theory and experiment}
\label{sec:comparison1}
As just explained at the end of the previous section, the average NMR line positions at $T <  T_{\mathrm{N}}$ (shown as black dots in Fig.~\ref{fig:Figure2}b) are independent of the contribution of the $L_6$ representation and reflect the influence of the LTSF and the uniform magnetization. This holds true even if a dipolar term is added to Eq.~\eqref{SHIFT1}, since the average NMR line position is not affected by the expected symmetrical dipolar splitting below $T_{\mathrm{N}}$.

With the external field in the $ab$ plane, the LTSF is always collinear with the $c$ axis.  For $T>T_{\mathrm{N}}$ the transverse magnetization induced by the LTSF contributes to $m_u$ in the form of Eq. \eqref{Feiguin_renormalized}. For $T<T_{\mathrm{N}}$ both $m$ and $m^i$ are still present. Since however $m$ (in Fig. \ref{fig:FigureCOMP}c) is weakly temperature dependent, the strong variation of the average NMR shift for $\mathbf{H} \parallel b$ at  $T<T_{\mathrm{N}}$ (see Fig.~\ref{fig:Figure5}c) is dominated by a contribution related to the $L_1(T)$ representation. 
Thus,  below $T_{\mathrm{N}}$, NMR allows us to reveal the effects of the interaction between the representations $L_1$ and $L_6$. We return to this issue after considering first the  $T>T_{\mathrm{N}}$ regime.

The relative shift captured in Eq.~\eqref{SHIFT1} includes the anisotropic orbital-shift tensor. Its components along the main crystal axes are usually determined via the so-called Clogston-Jaccarino plot, where the NMR line shift is plotted versus the corresponding susceptibility.\cite{Clogston61,Wolter05} This approach requires a sufficiently broad temperature range, in which the NMR shift mimics the sample's magnetization. Because of the large Cu-O exchange coupling in the \BACSO\ chains and the weak ${}^{29}$Si NMR signal at elevated temperatures, a reliable estimate of the orbital-shift components was not possible in this way. By assuming this tensor to be symmetric and temperature-independent, we released the parameters $\sigma_1$, $\sigma_2$ and $\sigma_5$ and the hyperfine couplings. In this way the whole data set could be fitted with a single set of parameters. The quantitative temperature dependences of $m$ and $m^i$ were established by using the calculations described in Sec.~\ref{sec:LTSF_physics}, inserting the values of the LTSF as obtained from the fits in Fig. \ref{fig:FigureCOMP}c. 

 In Fig.~\ref{fig:Figure5}a-d we display the result of the fits, as well as the individual contributions to the local magnetization, as a function of the angle $\theta$ and of temperature. In these figures the individual contributions to the total shift (red curve) of the NMR lines caused by the local longitudinal and transverse magnetization and by the orbital shift are highlighted as blue, black and green curves, respectively. For $\mu_0 H = 3.98$ T, only the global fit is presented. Also shown are the temperature independent contributions of the orbital shift (broken lines in Fig.~\ref{fig:Figure5}c and d). It may be seen that the temperature dependence of the shift due to the longitudinal magnetization is weak.
 The transverse component $|m_{\perp}|$ is small at 200 K but it grows significantly at low temperatures. The data were fitted in the range 20 to 230 K. The relative shift of the NMR lines (at very low fields with respect to saturation) scales linearly as a function of field.
 
 The fit parameters we obtain are $Y_1 \simeq 0.014$ T/$\mu_{\mathrm{B}}$, $Y_5 \simeq -0.16$ T/$\mu_{\mathrm{B}}$, $G_6 = -0.0752$ T/$\mu_{\mathrm{B}}$ and $G_3 = 0.1286$ T/$\mu_{\mathrm{B}}$. Since the fit parameters $Y_2$ and $Y_4$ are not lineary independent, we could fit only their combination $Y_2 g^{2,2} + Y_4 g^{1,1} = 0.53$ T/$\mu_{\mathrm{B}}$. The computed dipolar tensor components, to be inserted in Eq. \eqref{SHIFT1}, are of the order of 0.02--0.04 T/$\mu_{\mathrm{B}}$. The orbital shift values displayed in Fig.~\ref{fig:Figure5} are of the order of 150 ppm.
\subsection{\label{sec:compeSS} Competing spin structures}
Employing the same classification of representations as introduced in Ref.~\onlinecite{Glazkov04}, the antiferromagnetically-ordered phase in zero magnetic field is related to the $L_6$ representation and the corresponding order parameter. In the previous section we provided evidence for an enhanced transverse magnetic susceptibility even in the ordered regime. This enhancement is characteristic of quasi-1D chains in an  LTSF and we argue that it is the reason for the unusual spin-reorientation transitions that are observed in \BACSO. The microscopic approach requires considering the effects of the 1D-to-3D dimensional crossover in specific features of the magnetic properties. In case of chains with no LTSF, this was done with a combined mean-field and ``Random-Phase Approximation" approach.\cite{Schulz96} \\
Here we tackle the problem with a Ginzburg-Landau (GL) expansion\cite{Landau} of the free energy $\bar{\mathbf{\phi}}$ close to $T_{\mathrm{N}}$.
Although this phenomenological approach neglects fluctuation effects, it has the advantage of retaining the exchange-energy contributions to the susceptibility of the ordered phase. The mean-field approach also includes interactions between different, possibly coexisting,  order parameters. 

We start by constructing symmetry invariants of the little group of the $k$-vector.\cite{Balan17} In \BACSO , even if exposed to an applied field, a commensurate antiferromagnetic structure with $k=0$ is realized, leading to a little point group which coincides with $D_{2h}$. We call $L_{\beta\mu}$ the $\mu$-component of $\mathbf{L}_{\beta\mu}$, with $\mathbf{L}_{\beta}$ the $\beta$-IR; clearly the product $\beta \cdot \mu = 3N$ (with $N=8$ as the number of equivalent copper sites in the unit cell).
In terms of a GL free-energy expansion over all possible order parameters, a phase transition will occur whenever one of the coefficients $A_\beta$ of the quadratic term $A_\beta L^2_{\beta \mu}$ changes sign. Since the $L_{6c}$ representation is the one realized in the magnetically ordered regime in zero field, we can write that $A_6 = \varepsilon_6 (T-T_{\mathrm{N}})$ ($\varepsilon_6>0$). All the IRs $L_{\beta \mu}$ of the magnetic structures in \BACSO\ are one-dimensional.\cite{Glazkov04} It is therefore easy to construct invariant combinations of the $L_{\beta\mu}$ since the representations of the powers of these terms, which have to transform according to the trivial representation, remain one dimensional.
The expansion in Eq.~\eqref{VasiliyIRS} is based on the physics discussed in the previous sections of this paper.
Retained are the terms containing $L_{6\mu}$, related to the zero-field magnetic order, $L_{1\mu}$ representing the LTSF, and $H$ the external magnetic field. We will limit our considerations here to the case of an LTSF pattern $L_{1,c}$ which is realized for $\mathbf{H} \parallel b$. A more complete analysis will be published separately.\cite{Glazkov12}
The relevant expansion in powers of $L_{6\mu}$, $L_{1\mu}$ and $H_{\mu}$ reads:
\begin{align}
\bar{\mathbf{\phi}} &= \mathbf{\phi}_0 + A_6 \mathbf{L}^2_6 + A_1 \mathbf{L}^2_1 + B_6 \mathbf{L}^4_6 + B_{16} \mathbf{L}^2_6 \mathbf{L}^2_1 \nonumber \\
& + B'_{16} (\mathbf{L}_6 \cdot \mathbf{L}_1)^2 + D (\mathbf{H} \cdot \mathbf{L}_6)^2 + D' \mathbf{H}^2 \mathbf{L}^2_6 + a_a L^2_{6a} \nonumber \\
& + a_b L^2_{6b} + \alpha_b L_{1c}H_b + \alpha_c L_{1b}H_c - \frac{1}{2} \chi_p \mathbf{H}^2 - \frac{1}{2} \gamma_a H^2_a \nonumber \\
& - \frac{1}{2} \gamma_b H^2_b - \frac{\mathbf{H}^2}{8 \pi}, \label{VasiliyIRS}
\end{align}
with $\mathbf{\phi}_0$ as the value of the free energy in the paramagnetic phase in zero field.

The $A_1$ coefficient is positive above the N\'eel temperature, reflecting the absence of a spontaneous symmetry breaking related with the $\mathbf{L}_1$ order parameter. On the other hand we assume $A_1$ to be small  in the vicinity of $T_{\mathrm{N}}$ since, as indicated in Fig.~\ref{fig:Figure4}, $L_{1c}$ differs from the lowest energy-state configuration $L_{6c}$ only by the mutual orientation of the spins in the neighboring chains, which are relatively weakly coupled. The temperature dependence of $A_1$ is assumed to be linear in the vicinity of $T_{\mathrm{N}}$: $A_1 = A^{(0)}_1 \left[1+\varepsilon^{\mathrm{rel}}_1 (T-T_{\mathrm{N}})\right]$.

The fourth order term for $L_{6c}$ fixes the magnitude of the main order parameter below the transition; as required, $B_6 > 0$.
The terms with prefactors $B_{16}$ and $B'_{16}$ are crucial in our discussion, because they describe the exchange competition of the field-induced order $\mathbf{L}_{1}$ and the spontaneous order $\mathbf{L}_{6}$.  Microscopically, these terms arise from the simple idea that both the main order parameter and the induced order parameter involve the same local spins, eventually along the same crystallographic direction.
The $B_{16}$ coefficient is expected to be positive in order to enhance the energy cost for the coexistence of these two magnetic structures. Finally the term related to $B'_{16}$ defines the preferred mutual orientation of the two order parameters by means of the scalar product between the them. From the expansion in Eq.~\eqref{VasiliyIRS} alone it is not possible to predict whether a collinear ($B'_{16}<0$) or a transverse ($B'_{16}>0$) spin configuration is realized.

The terms related to $D$ and $D'$ describe interactions between the longitudinal magnetization and $L_6$. The terms with prefactors $a_{\mu}$  describe the orientation of the zero-field order parameter. Reported results of neutron diffraction\cite{Zhed02} and antiferromagnetic resonance\cite{Glazkov05} imply that $a_a>a_b>0$. The term $\alpha_b$ is responsible for the fact that the magnetic structure $L_{1c}$ is induced by an external field applied along the $b$ axis. The powers of $\mathbf{H}$ completing the expansion in Eq.~\eqref{VasiliyIRS} provide a full description of the effects of the $g$-tensor anisotropy on the longitudinal magnetization. This may be seen by recalling that:
\begin{equation}
\frac{\partial \bar{\mathbf{\phi}}}{ \partial \mathbf{H}} = - \frac{\mathbf{H}}{4 \pi} - \mathbf{M}. \label{magnLandau}
\end{equation}
Minimizing over the components of $\mathbf{L}_1$ for $\mathbf{H} \parallel b$ and assuming a zero-field collinear antiferromagnetism $L_{6c}$, we get $L_{1a} = L_{1b} = 0$ and:
\begin{align}
L_{1c} = \frac{- \alpha_b H_b}{2A_1(T) \left[ 1 + \frac{B_{16} + B'_{16}}{A_1(T)} L^2_{6c}(T) \right]}. \label{formulaL1a}
\end{align}
With a similar reasoning we obtain the longitudinal magnetization $M_b$ along the $b$-axis, in the former notation denoted as $m_u$:
\begin{align}
M_b = -2D (\mathbf{H} \cdot \mathbf{L}_6) L_{6c} -2D' \mathbf{L}^2_6 H_b - \alpha_b L_{1c} + (\chi_p + \gamma_b) H_b.
\label{formulaMa}
\end{align}
The last two equations deserve some discussion. 

Equation~\eqref{formulaL1a} captures the temperature dependence of the transverse staggered magnetization. Microscopically, the increase of $L_{1c}$ upon cooling above $T_{\mathrm{N}}$ ($L_{6c} = 0$) is due to the divergent transverse susceptibility of a 1D spin-\nicefrac{1}{2} quantum Heisenberg chain in an LTSF. 
This situation is modeled by the decrease of $A_1(T)$ on cooling (i.e.\ with $\varepsilon^{\mathrm{rel}}_1>0$). It also predicts a decrease of $L_1$ upon the growth of $L_6$ at $T<T_{\mathrm{N}}$, as observed in the NMR data. Microscopically, the change of regime upon decreasing temperature, from a divergent transverse susceptibility (characteristic of a spin chain) to a progressive competition between the field-induced magnetization pattern and the zero-field order parameter, is argued to be a direct consequence of the dimensional crossover from 1D to 3D of a chain in an LTSF. The clear experimental identification of how 1D physics affects the static magnetization properties even below $T_{\mathrm{N}}$ is the new result emerging from the present study. 
Below we address the question of how these anomalous properties for $T < T_{\mathrm{N}}$ can explain certain spin reorientation transitions observed in \BACSO.

\begin{figure}[b]
\includegraphics[width=0.5\textwidth]{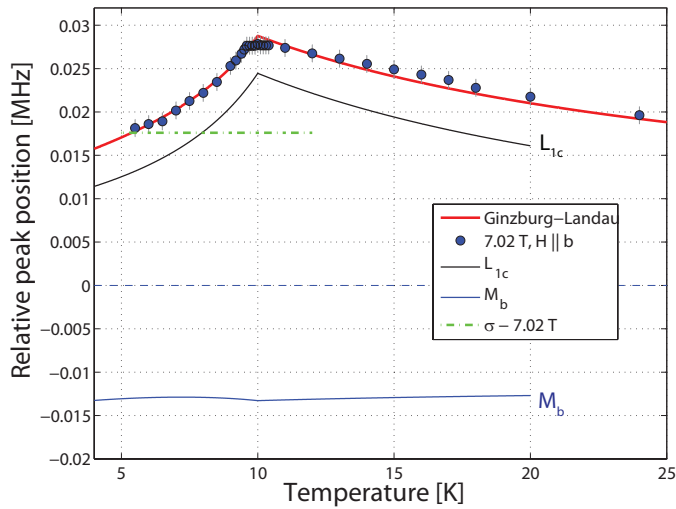} 
\caption{\label{fig:LandauFit} Data representing relative $^{29}$Si NMR line shifts in the vicinity of the ordering temperature, compared with the classical predictions of Eqs. \eqref{formulaL1a} and \eqref{formulaMa}. Although fully classical, the GL type approach grasps the competition between the spin structures described by $L_{1c}$ and $L_6$. The hyperfine parameters and the orbital shift are taken from the analysis for $T>T_{\mathrm{N}}$ (see text). The average NMR frequency is shown below the transition temperature. Black, blue and green line show single components of the fits due to $L_{1c}$, $M_b$ and $\sigma$, respectively.}
\end{figure}

By analyzing Eq.~\eqref{formulaMa} we note that the first term is zero for an easy axis ($c$ axis in our case) orthogonal to the direction of the applied field (along the $b$ direction). The second term, instead, provides corrections to the constant magnetization predicted by the standard mean-field theory below $T_{\mathrm{N}}$. Microscopically it can be related to a semi-classical contribution of spin-waves.\cite{Jongh10} 
From the magnetization data in Fig.~\ref{fig:FigureCOMP}c it may be concluded that $D' < 0$. Next we single out a constant paramagnetic contribution, with prefactors $\chi_p$ and $\gamma_b$, which is related to the magnetization of the ideal spin-\nicefrac{1}{2} Heisenberg chain at the N\'eel temperature. We note that also in this case a contribution from the staggered transverse susceptibility affects the longitudinal magnetization data, $M(T)$. While $L_{1c}$ scales as $\alpha_b H_b$, the contribution to $M_b$ scales as $\alpha^2_b H_b$, in full qualitative agreement with the microscopic approach. In magnetization measurements along either the $a$ or the $b$ axis, the contribution of the staggered magnetization matters, but it is not as outstanding as in NMR measurements, where both $M_b$ and $L_{c1}$ are revealed. 

The enhanced transverse susceptibility accounts very well for the observed spin reorientations. For example, with the applied field along the $b$-axis, such a transition occurs at $H_{\mathrm{sr}} \simeq 7.8$~T.\cite{Glazkov11} It is caused by a sudden change of the easy axis from the $c$- to the $a$-direction. Below $T_{\mathrm{N}}$ the quasi one-dimensionality 
extends itself in the form of a field-induced transverse susceptibility. It favours a field-induced spin alignment $L_{1c}$ which competes with the zero-field order parameter $L_{6c}$. The higher the field, the larger is the energy cost to sustain this arrangement [captured by the term $B_{16}$ of Eq.~\eqref{VasiliyIRS}]. Substituting Eq.~\eqref{formulaL1a} into \eqref{VasiliyIRS} yields an expression which depends on $\mathbf{L}_6$ only. A spin reorientation is then expected as the result of competition of the field-dependent anisotropic corrections with the conventional anisotropy of the order parameter at a field:
\begin{equation}
H_{sr} = \frac{2 A_1}{\alpha_b} \sqrt{\frac{a_a}{B'_{16}}}. \label{reorientField}
\end{equation}
It can be shown that, for $B'_{16}>0$, Eq.~\eqref{VasiliyIRS} also accounts for two spin-reorientation transitions when $\mathbf{H} \parallel c$. The inclusion of the staggered field pattern described by the representation $L_{2c}$ [see Eq.~\eqref{CombinationsLz}] could similarly account for the phase transition at $\mathbf{H} \parallel a$.\cite{Glazkov12}
Based on formulas \eqref{formulaL1a} and \eqref{formulaMa}, we now attempt a comparison with the experimental data in the temperature range $T<20$ K, where the  microscopic 1D model does not properly describe our results. 
Using the transferred hyperfine parameters determined in section \ref{sec:comparison1}, we compare the $^{29}$Si NMR line shift monitored for a field $H_{\mathrm{sr}}$ = 7.02~T oriented along the $b$-axis with the GL approach, postulating $L_{6c} = [(T_{\mathrm{N}}-T)/T_{\mathrm{N}}]^{\beta}$ and $\beta = 0.5$ (we arbitrarily set the zero-temperature limit of $L_{6c}$ to unity). This is shown in Fig.~\ref{fig:LandauFit}.
To obtain tentative estimates of the GL-model parameter, we first fitted $M_b$ of Fig.~\ref{fig:FigureCOMP}c in the vicinity of $T_{\mathrm{N}}$ to Eq.~\eqref{formulaMa} and obtained the parameters $\chi_p + \gamma_b \approx 7.05 \times 10^{-4}$ emu/mol Cu, $D' = -7.6 \times 10^{-5}$ emu/mol Cu and the ratio $\alpha^2_b/(2A^{(0)}_1) \approx 7.01 \times 10^{-4}$ emu/mol Cu. Notice that the present numbers contain already a prefactor 0.15 in the numerator of Eq.~\eqref{formulaL1a}, corresponding to the experimentally reported zero-temperature limit of $L_{6c}$ expressed in Bohr magnetons.
Considering the decrease of $L_{1c}$ with increasing temperature above $T_{\mathrm{N}}$ we obtain $\varepsilon^{\mathrm{rel}}_1 = $ 0.051 K$^{-1}$. Next, with the fixed $\alpha^2_b/(2A^{(0)}_1)$ ratio we could fit the relative NMR peak positions, as shown in Fig.~\ref{fig:LandauFit}, and hence determine the parameters $A^{(0)}_1$ and $B_{16}+B'_{16}$. The fit shown in Fig.~\ref{fig:LandauFit} was obtained with 
$A^{(0)}_1 \approx 3.15 \times 10^{7}$ emu/mol Cu and $B_{16}+B'_{16} \approx 7.63 \times 10^{7}$ emu/mol Cu, which correspond to a value $L_{1c}(T_{\mathrm{N}}) = 0.15 \cdot \alpha_b H_{b}/(2A^{(0)}_1) \approx 0.035$ $\mu_{\mathrm{B}}$ at 7.02~T.

\section{Summary and Conclusions}
\label{sec:summ}
A detailed analysis of $^{29}$Si NMR data obtained by probing single-crystalline \BACSO\ revealed the influence of 1D physics into the regime of 3D magnetic order at temperatures below 10 K. In this way the problem of weakly interacting nearest neighbor chains, described in the non-interacting limit by the model in Eq. \eqref{Hamorig}, could be addressed. Based on a classical Ginzburg-Landau analysis it is shown that in this type of compounds complicated ($H$, $T$) magnetic phase diagrams emerge. They are caused by the interaction of the transverse staggered local magnetization, originating from magnetic anisotropies in spin-\nicefrac{1}{2} Heisenberg chains, with the effective magnetic field due to the weakly ordered spin moments on neighboring chains. We argue that the previously established spin-reorientation transitions in \BACSO\ reflect this situation and can, therefore, be understood in this framework.

\begin{acknowledgments}
The authors thank K. Pr\v{s}a (EPF Lausanne) and O. Zaharko (PSI) for useful discussions. We are thankful to M. Zhitomirsky (CEA-Grenoble) for the enlightening comments concerning the Ginzburg-Landau approach. This work was financially supported in part by the Schweizerische Nationalfonds zur F\"{o}rderung der Wissenschaftlichen Forschung (SNF) and the NCCR research pool MaNEP of SNF. One of the authors (V.G.) thanks the Russian Foundation for Basic Research (RFBR) for the support of his studies.
\end{acknowledgments}

\end{document}